# Hydroxide-based magneto-ionics: electric-field control of reversible paramagnetic-to-ferromagnetic switch in α-Co(OH)$_2$ films

Alberto Quintana[1*], Abigail A. Firme[2], Christopher J. Jensen[1], Dongxing Zheng,[3] Chen Liu,[3] Xixiang Zhang,[3] and Kai Liu[1*]

[1]*Physics Department, Georgetown University, Washington, DC 20057, USA*
[2]*Department of Physics & Astronomy, University of Wyoming, Laramie, WY 82072, USA*
[3]*King Abdullah University of Science & Technology, Thuwal 23955-6900, Saudi Arabia*

## Abstract

Magneto-ionics has emerged as a promising approach to manipulate magnetic properties, not only by drastically reducing power consumption associated with electric current based devices but also by enabling novel functionalities. To date, magneto-ionics have been mostly explored in oxygen-based systems, while there is a surge of interests in alternative ionic systems. Here we demonstrate highly effective hydroxide-based magneto-ionics in electrodeposited α-Co(OH)$_2$ films. The α-Co(OH)$_2$, which is a room temperature paramagnet, is switched to ferromagnetic after electrolyte gating with a negative voltage. The system is fully, magnetically reversible upon positive voltage application. The origin of the reversible paramagnetic-to-ferromagnetic transition is attributed to the ionic diffusion of hydroxyl groups, promoting the formation of metallic cobalt ferromagnetic regions. Our findings demonstrate one of the lowest turn-on voltages reported for propylene carbonate gated experiments. By tuning the voltage magnitude and sample area we demonstrate that the speed of the induced ionic effect can be drastically enhanced.

*Corresponding Authors: Alberto Quintana (aq76@georgetown.edu), Kai Liu (kai.liu@georgetown.edu)

Traditionally, magnetic storage and spintronic devices are controlled using electric currents[1] which produce significant energy losses due to Joule heating. By replacing electric current with electric-field, the control of these magneto-electronic devices can be made to be more energy efficient. Among the different magneto-electric approaches to control magnetic properties,[2, 3] magneto-ionics, i.e. tuning magnetic response of a material by voltage-control of ion diffusion offers exciting potentials to manipulate material properties and enable a wide variety of magnetic functionalities,[4] thanks to its non-volatility, reversibility and tunability.[5] Importantly, magneto-ionics via electrochemical means is not hindered by the Thomas-Fermi screening length and thus has enabled magneto-electric effects in relatively thick films.[6]

So far, magneto-ionics has been mostly demonstrated in oxygen-based systems.[5, 7] Typically, those systems are composed of i) high oxygen mobility dielectrics – ferromagnetic (FM) metals[5, 7, 8] or ii) oxygen getters – oxides heterostructures.[9, 10] Magneto-ionics have allowed the tuning of a number of magnetic functionalities such as magnetization,[6] perpendicular magnetic anisotropy,[7, 11-13] coercivity,[14] domain wall velocity,[15] superconductivity,[16, 17] Dzyaloshinskii-Moriya interaction (DMI) and spin textures,[11, 18-20] ferrimagnetic order,[21] and exchange bias.[22, 23] There have also been recent interests in alterative ionic systems, e.g., based on hydrogen or nitrogen,[24, 25] that exhibit different ionic migration characteristics.

Electrolyte-gating, which exploits the occurrence of an electrical double layer to apply large electric fields (of about tenths of MV/cm),[26, 27] circumvents the necessity of thermal assistance thus achieving room temperature (RT) reversible oxygen-based magneto-ionics[6] without the use of oxygen buffer layers. In the last decade, electrolyte-gating has been established as a versatile approach to manipulate not only magnetism but also superconductivity, electrical conductivity and optical properties.[3, 28-31] However, the search for improved speed and lower voltage operation is still challenging.[32] Recently, magnetic switching in iron oxyhydroxide has shown promise through electrochemical redox reaction in aqueous media. However, once the magnetization is switched on, its paramagnetic state cannot be recovered and only a 15% tuning can be achieved.[33]

In this study, we have explored OH⁻ based magneto-ionics using alpha cobalt hydroxide [α-Co(OH)$_2$], as a potential candidate for low voltage and room temperature magneto-ionic actuation. The α-Co(OH)$_2$ phase belongs to the family of layered double hydroxides and is isostructural with hydrozincite-like compounds.[34] It is composed of positively charged sheets where Co$^{2+}$ is 6-fold coordinated with hydroxyl groups. However, partial protonation of hydroxyl ions resulted in charge unbalanced layers [Co$^{2+}$OH$_{2-x}$(H$_2$O)$_x$]$^{x+}$. The vacancies originated by the hydroxyl protonation are occupied by Co$^{2+}$ ions in tetrahedral coordination. Interlayer charge balance is achieved by the intercalation of polar or anionic species, e.g., water molecules, NO$_3^-$, CO$_3^{2-}$, Cl⁻ or organic surfactants, e.g., sodium dodecyl sulfate (SDS), among others.[35] As a result of the anionic intercalation, very large interlayer spacings are obtained. For example, when long organic molecules such as SDS are employed, interlayer spacings can exceed 1nm.[36] These structural features lead to outstanding pseudocapacitance properties thanks to an enhanced ionic conductivity.[37, 38]

In terms of magnetic behavior, α-Co(OH)$_2$ is reported to have a Néel-type antiferromagnetic alignment with an ordering temperature around 30 K. However, a more complex magnetic behavior has been reported due to a randomly averaged tetrahedra/octahedra arrangement which, especially for high tetrahedral fractions, results in uncompensated moments, glassy behaviors or FM-like hysteresis loops.[39] Despite the complex magnetic behaviors at low temperatures, α-Co(OH)$_2$ is a paramagnetic (PM) at room temperature.

Here we show that the PM α-Co(OH)$_2$ can be converted into FM under electrolyte-gating. By voltage polarity reversal, its PM character can be fully recovered. Since no room temperature FM-like character is expected for any of the cobalt hydroxides [α-Co(OH)$_2$ or β-Co(OH)$_2$], oxyhydroxide (CoOOH) or cobalt oxides (CoO and Co$_3$O$_4$), the origin of the observed phenomena can only be attributed to the reduction of PM α-Co(OH)$_2$ into FM Co. This is confirmed by X-ray diffraction and scanning transmission electron microscopy, where direct evidence of Co formation is observed. The response time, voltage and area dependences have been explored with the aim to improve the speed response of the induced FM signal.

**Experimental**

All the chemical reagents were of analytical grade and used as received: Cobalt nitrate hexahydrate (Co(NO$_3$)$_2$·6H$_2$O ACS reagent, ≥98%), Sodium dodecyl sulphate (SDS, NaC$_{12}$H$_{25}$SO$_4$, ACS reagent ≥99.0%), Ammonia nitrate (NH$_4$NO$_3$, ACS reagent, ≥98%), Sodium (Na, 99.9%), Propylene Carbonate anhydrous (C$_4$H$_6$O$_3$, 99.8%). Thin films of α-Co(OH)$_2$ were electrodeposited potentio-dynamically by sweeping potential from 0 down to -1 V (vs Ag$^+$/AgCl reference electrode) in a one-compartment 3 electrode cell, using a Princeton Applied Research EG&G 263A Potentiostat/Galvanostat.[40] Sputtered Ta (2 nm) / Pd (50 nm) films on Si (100) substrates were used as the working electrode, and a platinum spiral as the counter electrode. The electrolyte used for magneto-ionic samples contained 1 M cobalt nitrate, 0.35 mM SDS and 10 mM ammonium nitrate, with a pH value of 3.8. Electrolytes containing only 0.1 M and 1 M cobalt sulfate, respectively, were also used during sample optimization. Solutions were left unstirred during sample growth and all electrodes were placed vertically within the electrolyte. Sample thickness was tuned by the total number of potential sweeping cycles.

Sample morphology was studied using a Zeiss SUPRA55-VP scanning electron microscope (SEM). Crystal structure was analyzed by grazing incidence X-ray diffraction (GIXRD) in a Malvern-Panalytical X'Pert3 MRD diffraction system using Cu K$_\alpha$ radiation, an incidence angle of 0.4°, a step size of 0.02° and a linear detector with an integration time of 400 s in the 3°–52° 2θ range. Magnetic and magnetoelectric properties were measured using a Quantum Design MPMS3 superconducting quantum interference device (SQUID) magnetometer and a Princeton Measurements MicroMag3900 vibrating sample magnetometer. A home-made quartz cell for electrolyte-gating has been used, where a Pd foil was used as the counter electrode and the sample as the working electrode. Voltage was applied externally using an 2280S Keithley DC power supply. The monochromated aberration-corrected high-resolution scanning transmission electron microscopy (STEM, Titan 80-300, FEI) was performed to characterize the crystal structure of α-Co(OH)$_2$ film. The STEM samples were prepared by a focused ion beam (Helios 450, FEI).

**Results**

Samples of α-Co(OH)$_2$ were potentiodynamically grown from cobalt nitrate baths containing ammonia nitrate and sodium dodecyl sulphate by scanning the applied potential from 0V to -1V at a rate of 50mV/s. Further details from the sample synthesis and optimization are described in the Supplementary Information. A representative SEM image of an as-grown α-Co(OH)$_2$ sample is shown in **Fig. 1a**, illustrating a smooth and defect-free surface. Higher magnification image (**Fig. 1a inset**) reveals some degree of porosity, originated from the platelet-like morphology of α-Co(OH)$_2$.[41]

STEM has been used to examine the cross-section morphology of the as-grown sample, as shown in **Fig. 1b**. It confirms the fully compact α-Co(OH)$_2$ film as illustrated in Fig. 1a, thanks to the synergic effect of the higher cobalt nitrate concentration and the presence of the SDS (see Supplementary Information **Fig. S1**). High resolution images reveal the presence of crystalline planes with *d*-spacing of 0.6 nm (**Fig. 1c**) and 0.8 nm (**Fig. 1d**), which can be indexed with the (0012) and (009) basal planes, respectively, further confirming the intercalation of SDS.

Crystal structure of the electrodeposited films was studied by GIXRD. At high angles, the (104) reflection from an α-Co(OH)$_2$ with SDS intercalated is observed at $2\theta$=33.17°,[36] demonstrating a preferred crystalline orientation (**Fig. 2a**). The asymmetric "sawtooth" shape of the (104) peak is typical of turbostratic structures in 2-dimensional materials.[42] Here the [Co(OH)$_{2-x}$(H$_2$O)$_x$]$^{x+}$ layers have regular interlayer spacing, but deviate from perfect stacking ordering due to random translations or rotations along the normal direction. These defects lead to changes in interatomic distances, which are manifested in modifications of the peak position, intensity and peak broadening, and in turn the asymmetry of the peak shape.[43] At low angles, a small diffraction peak is observed at $2\theta$=4.2° (**Fig. 2b**) which belongs to the basal (003) reflection of α-Co(OH)$_2$. This corresponds to an interlayer spacing of about 2nm,[36, 44] evidencing the SDS intercalation during the electrodeposition process.[45] The origin of the relatively low intensity of (003) reflection can be attributed to the nanometric thickness of the grown platelets.[46]

In order to study the influence of electric fields on magnetic properties of the deposited films, the α-Co(OH)$_2$ films have been subjected to different voltage biases and gating times using an aprotic anhydrous electrolyte (propylene carbonate) with solvated sodium ions.[26] Electric fields were applied *ex-situ* in a homemade quartz cell. After voltage treatment, the samples were rinsed with isopropanol, dried, and immediately transferred into the SQUID. In-plane hysteresis loops were acquired for each gating condition, with the diamagnetic background subtracted. Room temperature magnetometry measurement of the as-grown (AG) sample is shown in **Fig. 3a** (black curve). The small signal and lack of hysteresis confirm the PM character of α-Co(OH)$_2$. After the application of -2 V for 120 min, a clear FM signal is observed (blue loop in **Fig. 3a**) with a coercivity of 330 Oe and saturation magnetization of about 2.5 emu·cm$^{-3}$. Here, magnetization values have been obtained by normalizing over the entire sample volume (area = 0.25cm$^2$ and thickness = 500μm). To the best of our knowledge, this is one of the lowest turn-on voltages obtained for propylene carbonate electrolyte-based experiments.[3] The FM hysteresis loop can be fully suppressed by simply reversing the applied voltage bias for the same gating duration (red curve in **Fig. 3a**), confirming the reversibility. A similar gating experiment was repeated at -4 V, shown in **Fig. 3b**. A clear hysteresis loop with a coercivity of 120 Oe can be observed after gating for 45 min, which is the time needed to obtain a similar saturation magnetization as that at -2 V for 120 min. The observed coercivity reduction may be ascribed to a more extensive ionic migration as a result of the larger applied voltage. This may promote more nucleation sites and therefore, resulting in the eventual percolation of these Co regions with in a lower coercive field.[25,47] The reversibility, i.e. recovering the PM character, is also demonstrated by a reverse gating (+4 V for 45 min). This means that the kinetics of the process can be sped up by a factor of 2.6 by simply increasing the gating voltage to -4 V without losing reversibility. Further evidence of magnetic reversibility is demonstrated by performing two consecutive gating cycles (±4V for 15 min), resulting in identical magnetization value for each cycle of about 0.7 emu·cm$^{-3}$ (see Supplementary Information Fig. S2).

The voltage dependence of the induced saturation magnetization ($M_s$) and its time evolution for different voltages (-2, -4, -6 and -8 V) were evaluated by measuring a series of hysteresis loops after different gating conditions, as shown in **Fig. 3c**. For all the tested voltages, $M_s$ scales linearly with time. As a reference point, a $M_s$ of ~2.5 emu·cm$^{-3}$ was achieved after a -2 V gating for 120 min. Interestingly we find that to achieve the same $M_s$, the gating duration is reduced to 10 min and 2 min (i.e. 12 and 60 times faster) at -6 V and -8 V, respectively. The dependence of induced $M_s$ on the applied voltage is clearly seen by defining a magnetoelectric parameter, $\alpha$, as the slope of the magnetization change over time for a constant gating voltage

$$\alpha = \left(\frac{dM_s}{dt}\right)_{V=constant}. \tag{1}$$

The exponential-like dependence of $\alpha$ with voltage (see **Fig. 3c inset**) has been previously reported in other ionic systems such as memresistors.[48, 49] From these results, an activation energy barrier of about 3-4 eV is estimated following calculations in previous studies.[12, 13, 50]

Additionally, the sample area dependence on the electrolyte-gating induced ferromagnetism has also been explored, as shown in **Fig. 3d**. It can be clearly seen that, the smaller the area, the larger the induced change for the same applied voltage and gating time. While the counter electrode area is preserved, the reduction of the sample area increases the charge density in the film and thus, the effective electric field. Therefore, increasing the bias and reducing the sample area can be exploited to achieve faster magneto-ionic responses.

In order to elucidate the origin of the observed magneto-ionic effect and any structural change involved, XRD patterns have been obtained in a sample treated at -4 V for 45 min, compared with another sample treated first at -4 V for 45 min and recovered at +4 V for a similar time (**Fig. 4**). The diffraction pattern of the as-grown film has been included as a reference, which exhibits the aforementioned SDS α-Co(OH)$_2$ (104) and Pd (111) peaks. For the -4 V voltage gated sample, a decrease in the (104) peak intensity is observed with no appreciable peak broadening. In addition, a minor reflection belonging to the (104) of an α-Co(OH)$_2$ with intercalated NO$^{3-}$ appeared around $2\theta \sim 36.0°$. Moreover, besides the α-Co(OH)$_2$ phase reflections, 2 sharp peaks from the β-Co(OH)$_2$ (100) and (011) diffractions have emerged at $2\theta \sim$ 32.6° and 38.8°, respectively. All the reflections from the deposited film exhibit a shift towards higher $2\theta$ angles with respect to either the as-grown sample or the literature values.[36] In contrast, the 2 peaks arising from the Pd buffer layer are shifted towards lower 2θ values, due to the formation of a stable PdH$_x$ phase.[51] Finally, after a second, positive +4V gating, the (104) α-Co(OH)$_2$ peak became much sharper, indicating significant grain growth. β-Co(OH)$_2$ peaks remain mostly unaffected with a new peak ascribed to the (002) reflection appearing. The diffraction peak positions are now in line with the literature values.

While the origin of the ferromagnetic signal upon electrolyte-gating can only be understood from the formation of ferromagnetic metallic Co regions (see Discussions below), for samples gated at -4V for 45min, the crystalline Co phase is below XRD detection limit. For illustration purpose, an as-grown α-Co(OH)$_2$ sample was gated for 3.5h at -8 V. As observed in Fig. 4, a clear (101) hcp Co peak is observed at ~ 47.3°, along with another less obvious shoulder at ~ 41.5°, partly overlapping with Pd (111), that can be indexed with the (100) hcp Co. Note that the formation of hcp Co, instead of fcc Co has been reported previously for electrolyte-gated systems.[52] Further evidence has been obtained by means of STEM characterization evidencing the formation of nanometer sized Co granules (Supplementary Information Fig. S3).

**Discussions**

The observed magneto-ionically induced transition from PM to FM state in the electrodeposited α-Co(OH)$_2$ films is a result of the electric-field induced ionic transport of OH$^-$ groups from the α-Co(OH)$_2$ phase, producing FM metallic cobalt. As shown in Fig. 3, the as-grown α-Co(OH)$_2$ is fully PM, excluding the presence of metallic cobalt or other FM impurities. Upon electrolyte-gating, only metallic Co can be responsible of the remarkable FM signal since none of the cobalt oxides/hydroxides/oxyhydroxides are FM at room temperature. From the magnetometry measurement shown in Fig. 3a, given the sample area, it can be estimated that the corresponding net Co thickness would be ~ 1nm, which makes it difficult to detect directly via XRD. Increasing the Co thickness by gating at larger bias for longer times has allowed the direct detection of hcp-Co by X-ray diffraction (Fig.4) and STEM (Supplementary Information Fig. S3). The broad Co peak detected in XRD is in agreement with the nanometric character of Co formed by electrolyte-gating.[6]

Besides pure electric-field induced ionic migration, some electrochemical reactions may potentially contribute to the observed phenomena, but can be excluded as discussed below. For example, the Co formation may be produced by proton ($H^+$) mediated electrochemical reduction. However, due to the anhydrous character of the chosen electrolyte this can be discarded.[24] Moreover, $Li^+$ or $Na^+$ may also promote the reduction of cobalt species into metallic cobalt.[53] However, the very low concentration of $Na^+$ in our electrolyte, which is about few ppm[54], rules out this possibility. Finally, it is reasonable to consider that the organic electrolyte may play a role, especially at such overvoltage. The oxidation of propylene has been reported to generate $CO_2$ and propylene oxide, the former released as gas and the latter remaining in the solution,[55] thus suggesting that the electrolyte is not mediating the reduction. Importantly, the decomposition of the electrolyte can be suppressed by combining propylene carbonate with other organic solvents such as dimethyl carbonate or methylethyl carbonate.[56]

Further evidence of ionic transport is obtained from the magnetization dependence with time and the different applied voltages. The α defined in Eqn. 1 presents an exponential dependence of induced magnetization with the magnitude of the applied voltage (Fig. 3c inset). This dependence can be understood by the hopping mechanism of Mott and Gurney, which states that the ions jump between adjacent sites surpassing a thermally activated energy barrier.[57] For high enough electric fields, as those produced by electrical double layers, the ionic transport depends exponentially on the applied electric field,[57]

$$j \sim e^{\left(\frac{-w_a^0}{kT}\right)} e^{\left(\frac{aze_0E}{2kT}\right)}, \qquad (2)$$

where $j$ is the ionic current density, $e_0$ the electron charge, $z$ the atomic number, $a$ the distance between adjacent sites, $W_a^0$ the energy barrier, $k$ the Boltzmann constant, $T$ the temperature and $E$ the applied electric field. When a sufficiently large electric field (> the turn-on voltage) is applied to the α-Co(OH)$_2$ film, OH$^-$ starts to move, creating FM regions composed of metallic cobalt. While we cannot directly measure this ionic motion, the amount of the resultant FM regions can be measured. Once the electrical double layer is steady, the ionic motion of OH$^-$ is described by $j$, as shown in Eqn. 2, which is constant over time. As $j$ is linked to the size of the FM region created, this means that the induced magnetization is linear over time, in agreement with results shown in Fig. 3c. If the applied voltage increases, $j$ will increase accordingly and therefore the rate of the induced magnetization, i.e. the slope of magnetization vs time, will increase as well. This implies that $\alpha$ is directly related to $j$ and therefore, the exponential dependence of $j$ with the external voltage must be preserved, as shown by the inset of Fig. 3c.

The ionic motion of OH$^-$ across a double layered hydroxide, like the α-Co(OH)$_2$, has been described by a Grotthuss mechanism, which depicts the OH$^-$ motion by the hopping of the OH$^-$ via fast creation/dissociation of hydrogen bonds.[58] Recently, it has been reported that indeed, double layered hydroxides exhibit the coexistence of both diffusion and Grotthuss-like ionic motion.[59] Another telltale sign is the activation energy, which has been reported to be on the order of 0.2 – 0.3eV for Grotthuss motion.[60] In our case, using Eqn. 2,[12, 50] the energy barrier is determined to be between 3-4 eV, implying the motion is governed by diffusion rather than the Grotthuss motion.

Upon negative gating, the decrease in the intensity in the (104) reflection is consistent with a reduction of the overall α-Co(OH)$_2$ content, suggesting that ionic transport has occurred. No appreciable peak broadening is observed, indicating that the induced ionic transport does not promote significant grain growth, as reported in oxide systems[6]. Charge balance in Co(OH)$_2$ layers due to the presence of tetrahedrally coordinated Co$^{2+}$ ions, and therefore its stability, is assisted by the intercalation of water molecules or other species such as NO$^{3-}$ or SDS molecules.[35, 37] It has been reported that upon contact with hydroxide solutions, the deintercalation of these stabilizing molecules promotes the formation of β-Co(OH)$_2$.[61] Thus, the appearance of new sharp reflections from the β-Co(OH)$_2$ suggests that the negative voltage application has triggered the deintercalation of those stabilizing molecules. The conversion of α-Co(OH)$_2$ into β-Co(OH)$_2$ takes place through the Oswald ripening process, i.e. dissolution and regrowth, thus allowing the formation of larger crystallites, i.e. sharper diffraction peaks[61, 62] as shown in Fig. 4.

Finally, both α-Co(OH)$_2$ and β-Co(OH)$_2$ peaks on the gated samples are observed to shift towards higher 2$\theta$ angles. Taviot-Guého *et al.* reported that upon electro-oxidation, layered-double hydroxides (LDH), such as Co-Al LDH shift towards higher 2$\theta$ due to the deintercalation of stabilizing ions and due to the oxidation in KOH solutions of Co$^{2+}$ ions into Co$^{3+}$ which possess a smaller ionic radius[63]. However, the induced structural changes upon electro-oxidation are irreversible and no shift towards the initial position would occur, while our sample reverses back upon positive bias application. Moreover, oxidation processes are not consistent with the applied voltage polarity in our experiments. One possible explanation of the observed peak shift can be extracted by analyzing the buffer layer peaks. While the as-grown sample exhibits diffraction peaks consistent with Pd (JDPDS 46-1043), upon negative bias application, those peaks shift towards lower $2\theta$ angles, consistent with the formation of β-PdH$_x$.[64] Hydrogenation of Pd leads to a drastic lattice parameter increase[51], and thus transferring a large strain to the hydroxide layer grown on top, shifting both α and β- Co(OH)$_2$ peak positions. The origin of the palladium hydride phase formation can be attributed to the splitting of the intercalated water molecules in the α-Co(OH)$_2$ phase, which upon voltage application, dissociate at the Pd interface following the hydrogen evolution reaction.[65] The produced hydrogen is stored in the Pd electrode as a PdH$_x$ phase and the OH$^-$ can be either released into the electrolyte[29, 66] or stabilized in the adjacent hydroxyl layers. Interestingly, this shows the potential use of LDH as a reservoir for H$_2$-based magneto-ionics[24].

**Conclusions**

We have demonstrated effective room-temperature hydroxide-based magneto-ionic PM to FM switch in α-Co(OH)$_2$. The use of this layered double hydroxide results in one of the lowest turn-on voltages, -2V, reported so far in propylene carbonate-based electrolyte gating experiments. When subjected to a negative gating, ferromagnetism is induced in the originally PM α-Co(OH)$_2$, accompanied by a reduction in the (104) peak intensity. The process is attributed to an electric-field induced ionic migration of OH$^-$ groups from the α-Co(OH)$_2$ phase into the electrolyte, producing FM metallic Co regions. When a positive bias is applied, the initial PM state is recovered. Structurally, the intensity of the (104) reflection is recovered but the loss of the asymmetry and the sharpness of the diffraction peaks reveals that the reversal produces a better crystallized and more ordered α-Co(OH)$_2$. Here it is also demonstrated that increasing the gating voltage up to 8V, the process is sped up by factor 60. It is shown that the decrease in the sample area also speeds up the process due to a larger charge accumulation at the sample and, therefore, a more intense electric-field.

**Acknowledgements**

This work has been supported in part by SMART (2018-NE-2861), one of seven centers of nCORE, a Semiconductor Research Corporation program, sponsored by National Institute of Standards and Technology (NIST), the NSF (ECCS-1933527, ECCS-2151809), and KAUST (OSR-2019-CRG8-4081). A.A.F. acknowledges support from the NSF-REU program (DMR-1659532). The acquisition of a Magnetic Property Measurements System (MPMS3), which was used in this investigation was supported by the NSF-MRI program (DMR-1828420).

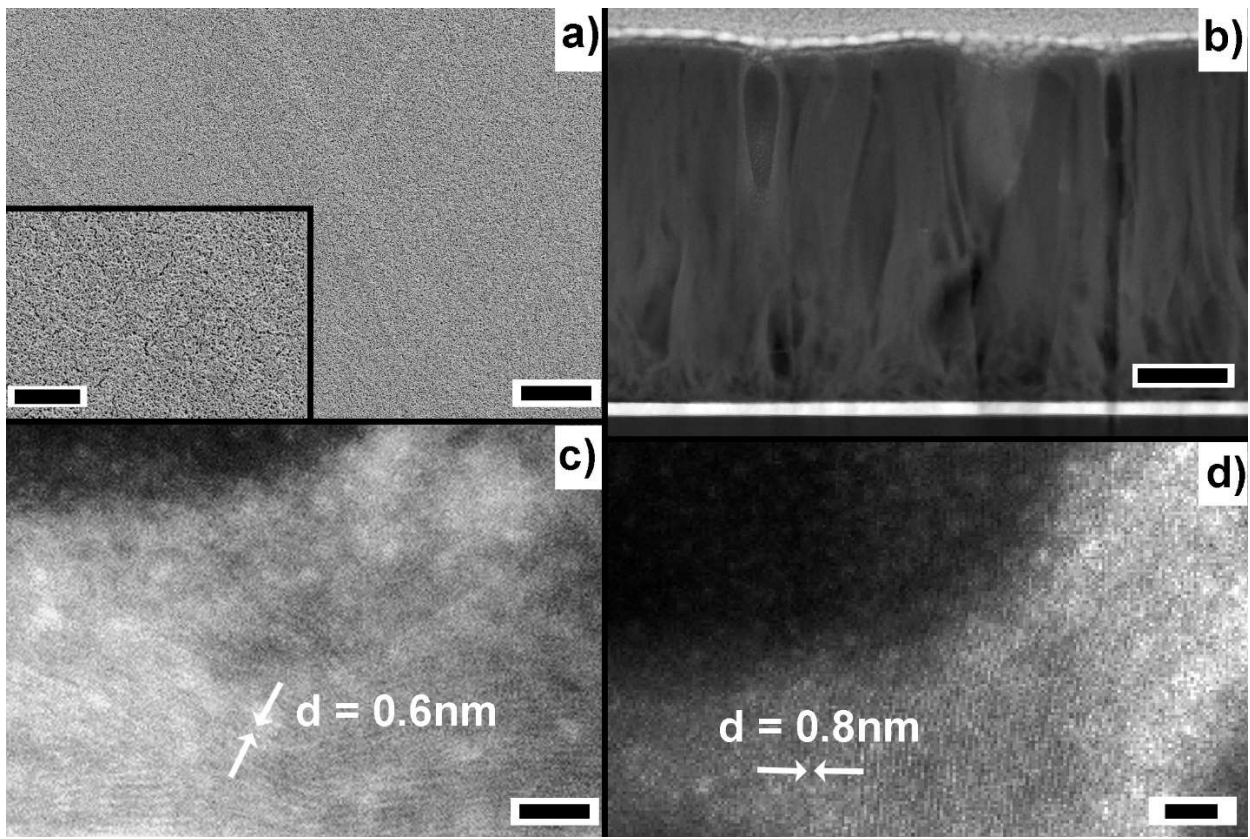

**Fig. 1**: a) Low and high resolution (inset) SEM images of the synthesized α-Co(OH)$_2$. In b) low magnification STEM image of the cross-section view of an as-grown sample and, in c) and d), its corresponding high-resolution images. Scale bar is 10 μm in image (a) and 4 μm in its inset, 150 nm in (b), and 5 nm in (c-d).

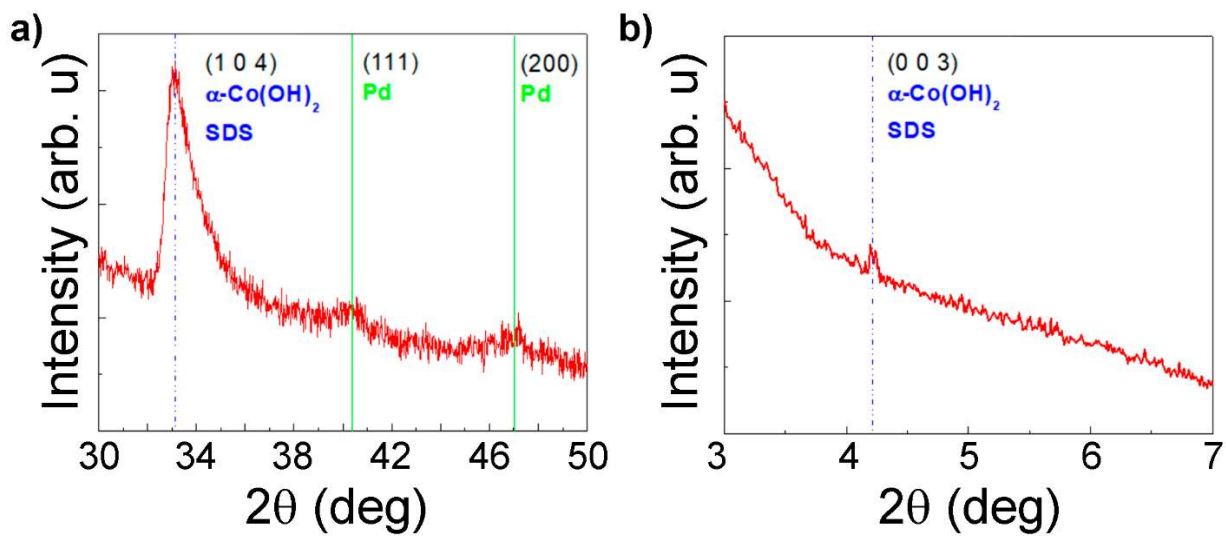

**Fig. 2**: Grazing incidence X-ray diffraction pattern of the as-grown α-Co(OH)$_2$, in **a)** high-angle and **b)** low angle regions. Dashed blue line identifies α-Co(OH)$_2$ with intercalated SDS and green lines identify peaks from the Pd buffer layer.

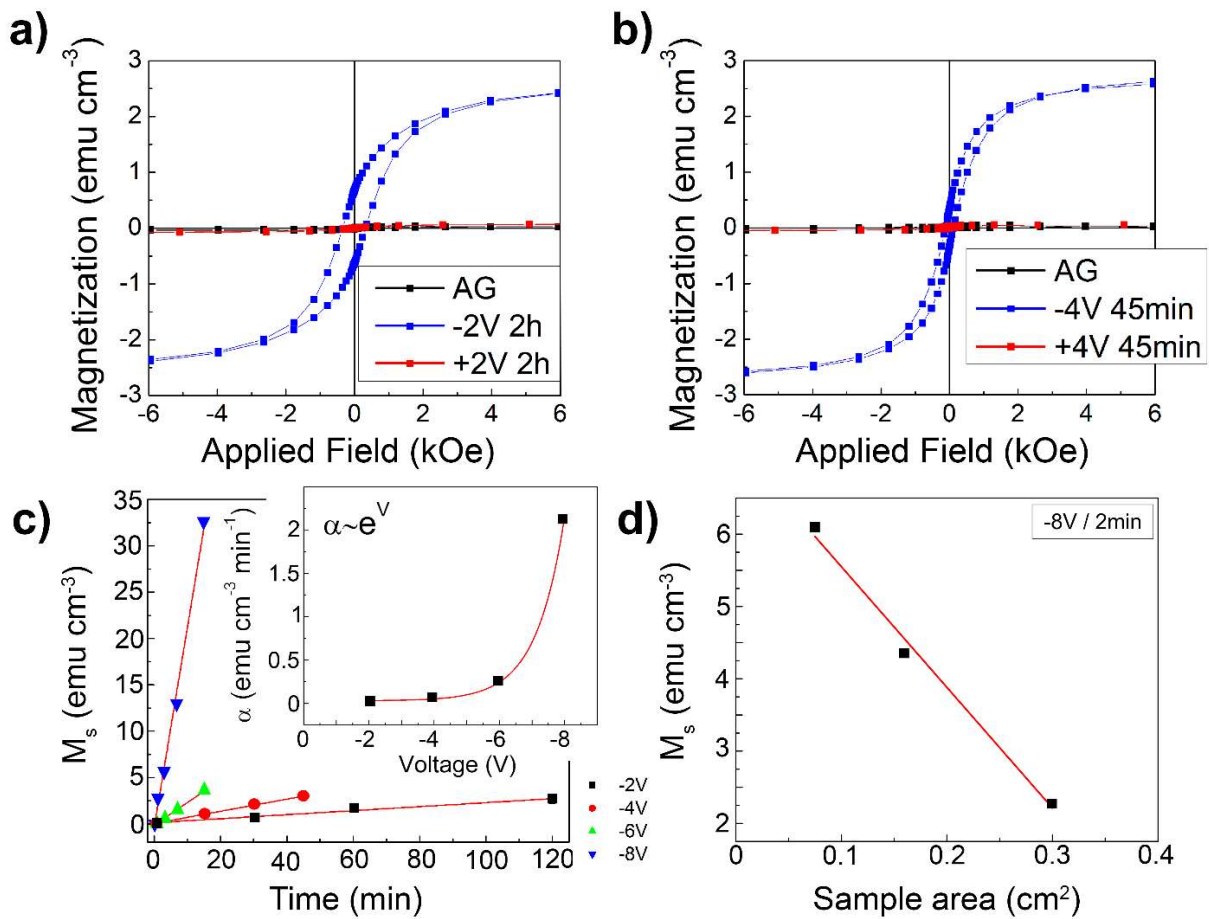

**Fig. 3**: **a)** Room temperature hysteresis loops for the as-grown, -2 V (120 min) and +2 V (120 min) electrolyte-gated samples. **b)** Room temperature hysteresis loops for the as-grown, -4 V (45 min) and +4 V (45 min) treated samples. **c)** Time dependence of the magnetization for different gating voltages. Inset shows the dependence of $\alpha$ on the gating voltage. **d)** Sample area dependence of the magnetization for a fix gating voltage and time.

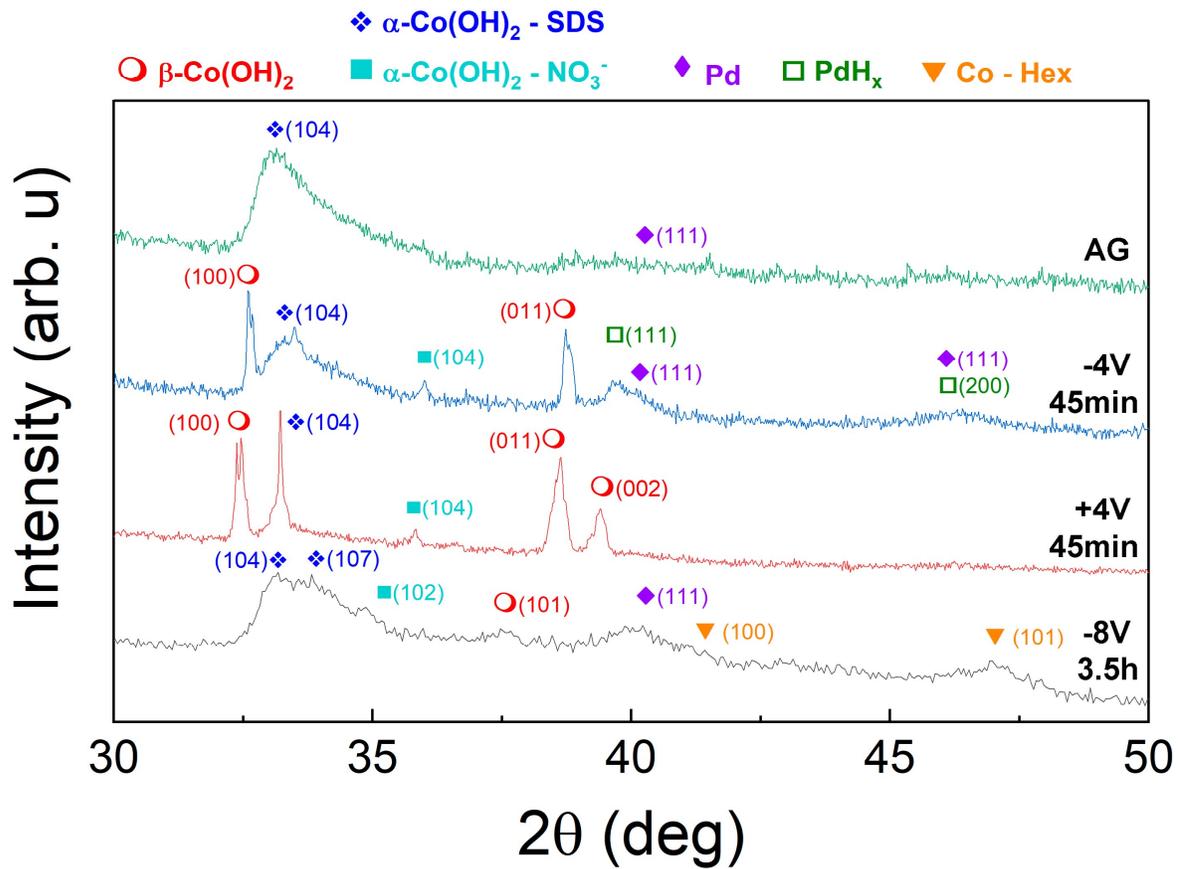

**Fig. 4**: X-ray diffraction pattern of 1) as-grown sample, 2) gated sample at -4 V for 45 min, 3) gated sample at -4 V for 45 min and recovered at +4 V for 45 min, and 4) gated sample at -8V for 3.5 h.

# Supplementary Information

# Hydroxide-based magneto-ionics: electric-field control of reversible paramagnetic-to-ferromagnetic switch in α-Co(OH)$_2$ films

Alberto Quintana[1*], Abigail A. Firme[2], Christopher J. Jensen[1], Dongxing Zheng,[3] Chen Liu,[3] Xixiang Zhang,[3] and Kai Liu[1*]

[1]Physics Department, Georgetown University, Washington, DC 20057, USA

[2]Department of Physics & Astronomy, University of Wyoming, Laramie, WY 82072, USA

[3]King Abdullah University of Science & Technology, Thuwal 23955-6900, Saudi Arabia

*Corresponding Authors: Alberto Quintana (aq76@georgetown.edu), Kai Liu (kai.liu@georgetown.edu)

# Supplementary Information

1. **Sample Synthesis**

The electrodeposition of cobalt hydroxide from cobalt nitrate bath is originated due to a local pH increase around the working electrode when nitrate ions are reduced, according to the following two possible chemical reactions:

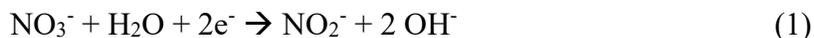

$$NO_3^- + H_2O + 2e^- \rightarrow NO_2^- + 2\ OH^- \qquad (1)$$

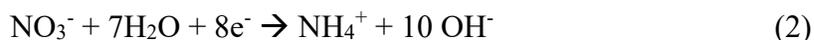

$$NO_3^- + 7H_2O + 8e^- \rightarrow NH_4^+ + 10\ OH^- \qquad (2)$$

This reduction occurs at lower overpotentials than the metal Co deposition, producing an increase of $OH^-$ groups in the vicinity of the working electrode, hindering the cobalt metal deposition and promoting the formation of cobalt hydroxide. In some cases, especially when acidic electrolytes are employed, metallic cobalt co-deposition within the hydroxide can occur. It has been reported that the ratio of cation/nitrate anion plays an important role. Thus, to assure the cobalt deposition suppression, ammonia nitrate ($NH_4NO_3$) is introduced. This is crucial to achieve fully paramagnetic samples.[1]

As a first attempt, α-$Co(OH)_2$ films were electrodeposited using electrolytes with a $Co(NO_3)_2 \cdot 6H_2O$ concentration of 0.1M.[2] Their morphology consist of hexagonal platelets randomly oriented with the basal planes perpendicular to the substrate surface (**Fig. S1a**).[3-4] This morphology leads to films with a large surface-to-volume ratio (i.e. porosity). Even though certain porosity could be beneficial, such as enhancing the magnetoelectric effects,[5-7] these films present uneven surfaces due to size dispersion of platelets. For this reason, their roughness needs to be minimized for thin-film based device applications.

# Supplementary Information

When Co(NO$_3$)$_2$·6H$_2$O concentration in the electrolyte was increased to 1 M, the sample morphology improved with a more uniform surface (**Fig. S1b**). Close examination of the films still revealed a certain degree of porosity from the perpendicular growth of the platelets, but the films were more densely packed (**Fig. S1b inset**). However, cracks appeared all over the surface, produced by stress relaxation during sample growth.[8] When SDS was incorporated to the electrolyte, the film morphology was preserved and, importantly, crack formation was suppressed (**Fig. 1a** in main text). It has been reported that SDS acts as a stress-relieve agent during electrodeposition, avoiding crack formation.[8-9]

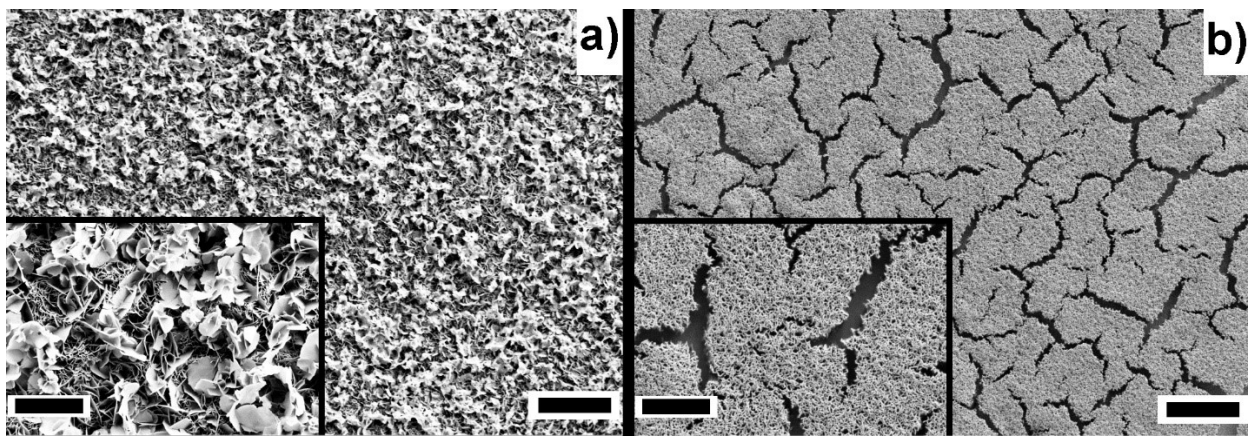

**Fig. S1:** Low and high resolution (inset) SEM images of the samples synthesized from an electrolyte containing a) 0.1 M and b) 1 M Co(NO$_3$)$_2$·6H$_2$O. Scale bar is 10 μm in images (a-b), and 4 μm in (a-b) insets.

# Supplementary Information

2. <u>**Magnetic Switching**</u>

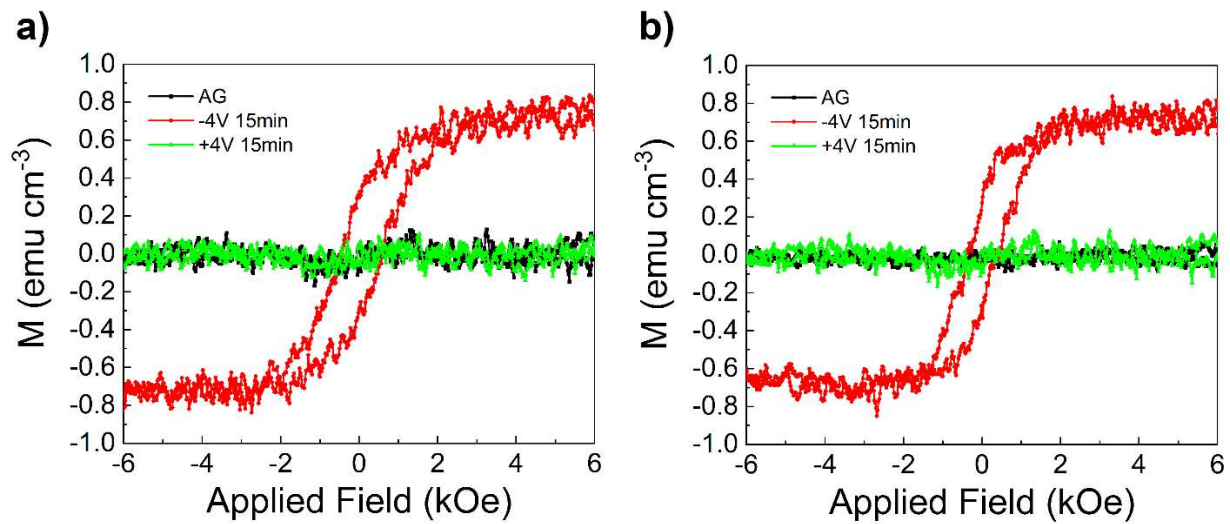

**Fig. S2:** Magnetic hysteresis loops for two consecutive switching cycles in α-Co(OH)$_2$ films after gating it at -4V and +4V for 15min each for a) one cycle and b) two cycles, respectively. Note that the AG in panel b) corresponds to the +4V 15 min state from panel a).

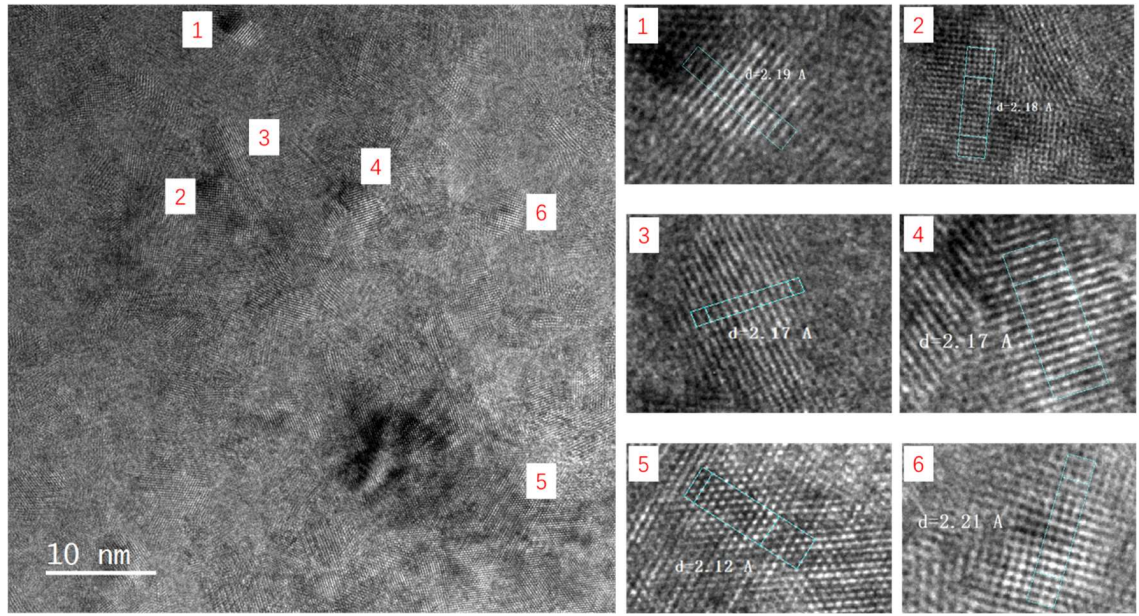

**Fig. S3:** High resolution scanning transmission electron microscopy and zoom-in images of labeled regions collected from an α-Co(OH)$_2$ sample after a gating procedure of -8V for 3.5h. The *d*-spacing corresponds to (100) hcp Co.

# Supplementary Information